\documentclass[version1996,10pt]{siggraph}%{article}
\usepackage{epsf}
\usepackage{graphicx}
\usepackage{wrapfig}
\usepackage{amssymb}
\usepackage{amsmath}
\newlength{\mywidth}\mywidth=2.3truein %width of caption
\epsfysize=\mywidth
\epsfxsize=\mywidth
\renewenvironment{figure}{\refstepcounter{figure}
\baselineskip=0.4\normalbaselineskip\footnotesize}
{\baselineskip=\normalbaselineskip}
\def\caption{{\bf Fig.\arabic{figure}.\quad}}

\begin{document}

%+Title
\title{Exploring Oracle\textregistered  \ RDBMS latches  using Solaris\texttrademark  DTrace}
\author{Andrey Nikolaev RDTEX LTD, Russia,
\\ Andrey.Nikolaev@rdtex.ru,
\\ \texttt{http://andreynikolaev.wordpress.com}
\\ {Proceedings of MEDIAS 2011 Conference. 10-14 May 2011. Limassol, Cyprus. ISBN 978-5-88835-032-4} }
%\date{\today}
%\keywords{Oracle, Spinlock, Latch, DTrace, Spin Time, Spin-Blocking}
\maketitle
\setcounter{page}{135}

\section*{Abstract}
{\it  Rise of hundreds cores technologies bring again to the first plan the problem of interprocess synchronization in database engines. Spinlocks are widely used in contemporary DBMS to synchronize processes at microsecond timescale. Latches are Oracle\textregistered\   RDBMS specific spinlocks.  The latch contention is common to observe in contemporary high concurrency OLTP environments.

In contrast to system spinlocks used in operating systems kernels, latches work in user context. Such user level spinlocks are influenced by context preemption and multitasking. Until recently there were no direct methods to measure effectiveness of user spinlocks. This became possible with the emergence of Solaris\texttrademark  10 Dynamic Tracing framework. DTrace allows tracing and profiling both OS and user applications.

This work investigates the possibilities to diagnose and tune Oracle  latches.
It explores the contemporary latch realization and spinning-blocking strategies, analyses corresponding statistic counters. 

A mathematical model developed to  estimate analytically the effect of tuning  \textbf{\_SPIN\_COUNT} value.
%\end{abstract}
%\keywordlist
}

{\bf Keywords:}  Oracle, Spinlock, Latch, DTrace, Spin Time, Spin-Blocking

\section{Introduction}
According to latest Oracle\textregistered\   documentation \cite{Concepts112} latch is ''A simple, low-level serialization mechanism to protect shared data structures in the System Global Area''. 

Huge OLTP  Oracle  RDBMS ''dedicated architecture'' instance  contains  thousands  processes  accessed the shared memory. This shared memory is called ''System Global Area'' (SGA) and consist of millions cache, metadata and result structures. Simultaneous processes access to these structures synchronized by Locks, Latches and KGX Mutexes:

\begin{figure}
\begin{center}
\includegraphics[width=0.5\textwidth]{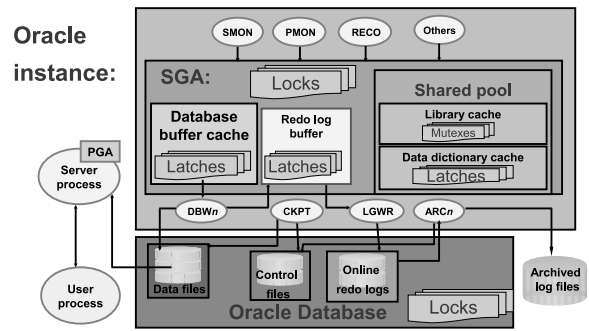}

\vspace*{-10mm}
\caption{Oracle\textregistered\  RDBMS architecture}\label{fig:arc}
\end{center}
\end{figure}

 Latches and KGX mutexes are the Oracle   realizations of general spin-blocking spinlock concept. The goal of this work is to explore  the  most commonly used spinlock inside Oracle --- latches.  Mutexes appeared in latest Oracle versions inside Library Cache only. Table \ref{tab:scomp} compares these synchronization mechanisms.
% {\em{}

\textbf{Wikipedia} defines the spinlock as {\em{}''a lock where the thread simply waits in a loop (''spins'') repeatedly checking until the lock becomes available. As the thread remains active but isn't performing a useful task, the use of such a lock is a kind of busy waiting''}.

Use of spinlocks for multiprocessor synchronization were first introduced by Edsger Dijkstra in \cite{Dijkstra65}. Since that time, a lot of researches were done in the field of mutual exclusion algorithms. Various sophisticated spinlock realizations were proposed and evaluated. The contemporary review of such algorithms may be found in \cite{Anderson2003}

There exist two general spinlock types:
\begin{itemize}
\item   System spinlock. Kernel OS threads cannot block. Major metrics to optimize system spinlocks are atomic operations (or Remote Memory References) frequency and shared bus utilization. 
\item   User application spinlocks like Oracle latch and mutex. It is more efficient to poll the latch for several  usec rather than pre-empt the thread doing 1 ms context switch. Metrics are latch acquisition CPU and elapsed times.
\end{itemize}

The latch is a hybrid user level spinlock. The documentation named subsequent latch acquisition phases as:
\begin{itemize}
\item Atomic \textbf{Immediate Get}.
\item If missed, latch spins by polling location nonatomically during \textbf{Spin Get}.
\item In spin get not succeed, latch sleeps for \textbf{Wait Get}.
\end{itemize}

According to Anderson classification \cite{Anderson90}  the latch spin is one of the simplest spinlocks -√ TTS ({\em{}''test-and-test-and-set''}). %It polls the location  nonatomically and uses atomic instructions to finally acquire the latch.
 
 Frequently spinlocks  use more complex structures than TTS.  Such algorithms, like famous MCS spinlocks \cite{MCS} were designed and benchmarked to work in the conditions of ~100\% latch utilization and may be heavily affected by OS preemption. For the current state of spinlock theory see \cite{Art08}.

If user spinlocks are holding for long, for example due to OS preemption, pure spinning becomes ineffective. To overcome this problem, after predefined  number of spin cycles latch waits (blocks) in a queue.  Such spin-blocking was first introduced in \cite{oust82} to achieve balance between CPU time lost to spinning and context switch overhead. Optimal strategies how long to spin before blocking were explored in \cite{agarwal91,karlin90,boguslavsky91}. Robustness of spin-blocking  in contemporary environments  was recently investigated in \cite{look2009} 

Contemporary servers having  hundreds CPU cores  bring  to the first plan the problem of spinlock SMP scalability. Spinlock utilization increases almost linearly with number of processors \cite{sinharoy96}.  One percent spinlock utilization of Dual Core development computer is negligible and may be easily overlooked. However, it may scales upto 50\% on 96 cores production server and completely hang the 256 core machine. This phenomenon  is also known as "Software lockout". 

\hfill Table \ref{tab:scomp}. {Serialization mechanisms in Oracle}
\vspace*{-5mm}
\begin{center}\label{tab:scomp}
\small{\begin{tabular}{|p{0.25\linewidth}|p{0.15\linewidth}|p{0.23\linewidth}|p{0.18\linewidth}|}
\hline
&\textbf{Locks}&\textbf{Latches}&\textbf{Mutexes}\\
\hline
\textbf{Access}&Several Modes&Types and Modes&Operations\\
\hline\textbf{Acquisition}&FIFO&SIRO (spin) + FIFO&SIRO \\
\hline\textbf{SMP Atomicity}&No&Yes&Yes\\
\hline\textbf{Timescale}&Milli\-seconds&Micro\-seconds&SubMicro\-seconds\\
\hline\textbf{Life cycle}&Dynamic&Static&Dynamic\\
\hline
\end{tabular}
}
\end{center}
\normalsize

%\begin{figure}
%\begin{center}
%\includegraphics[width=0.5\textwidth]{lcontention.eps}
%\caption{Oracle instance hangs due to  heavy \textit{''cache buffers chains''} %latch contention}\label{fig:lcont}
%\end{center}
%\end{figure}

\subsection{Oracle\textregistered\  RDBMS  Performance Tuning overview}

During the last 30 years, Oracle developed from the first tiny one-user SQL database to the most advanced contemporary RDBMS engine. Each version introduced new performance and concurrency advances. The following timeline is the excerpt from  this evolution:
{\small 
\begin{itemize}
\item v. 2 (1979): the first commercial SQL RDBMS
\item v. 3 (1983): the first database to support SMP 
\item v. 4 (1984): read-consistency, Database Buffer Cache 
\item v. 5 (1986): Client-Server, Clustering, Distributing Database, SGA
\item v. 6 (1988): procedural language (PL/SQL), undo/redo, \textbf{latches}
\item v. 7 (1992): Library Cache, Shared SQL, Stored procedures, 64bit
\item v. 8/8i (1999): Object types, Java, XML
\item v. 9i (2000): Dynamic SGA, Real Application Clusters
\item v. 10g (2003): Enterprise Grid Computing, Self-Tuning, mutexes
\item v. 11g (2008): Results Cache, SQL Plan Management, Exadata 
\item v. 12c (2011): Cloud. Not yet released at the time of writing 
\end{itemize}
}

As of now, Oracle\textregistered\   is the most complex and widely used SQL RDBMS. However, quick search finds more then 100 books devoted to Oracle performance tuning on Amazon \cite{adams,millsap,OWI}. Dozens conferences covered this topic every year. Why Oracle needs such tuning?

Main reason of this is complex and variable workloads. Oracle is working in so different environments ranging from huge OLTPs, petabyte OLAPs to hundreds of tiny instances running on one server. Every database has its unique features,  concurrency and scalability issues.

To provide Oracle RDBMS ability to work in such diverse environments, it has  complex internals.  Last Oracle version 11.2 has 344 ''Standard''  and  2665 ''Hidden'' tunable parameters to adjust and customize its behavior. Database administrator's education is crucial to adjust these parameters correctly.

Working at Support, I cannot underestimate the importance of developer's education. During design phases, developers need to make complicated algorithmic, physical database and schema design decisions. Design mistakes and ''temporary'' workarounds may results in million dollars losses in production. Many ''Database Independence'' tricks also results in performance problems.

Another flavor of performance problems come from self-tuning and SQL plan instabilities, OS and Hardware issues. One need take into account also more than 10 million bug reports on MyOracleSupport. It is crucial to  diagnose the bug correctly.

Historically latch contention issues were hard to diagnose and resolve.  Support engineers definitely need more mainstream science support. This work summarizes author investigations in this field.

To allow diagnostics of performance problems Oracle instrumented his software well.
Every Oracle session keeps many statistics counters. These counters describe {\em{}''what sessions have done''}. There are 628 statistics in 11.2.0.2.

 Oracle Wait Interface events complements the statistics. This instrumentation describes  {\em{}''why Oracle sessions have waited''}. Latest 11.2.0.2 version of Oracle accounts 1142 distinct wait events. Statistics and Wait Interface data used by Oracle\textregistered\   AWR/ASH/ADDM tools, Tuning Advisors, MyOracleSupport diagnostics and tuning tools.  More than 2000 internal ''dynamic performance'' \textbf{X\$} tables provide additional data for diagnostics. Oracle performance data  are  visualized by  Oracle Enterprise Manager and other specialized tools.
 
This is the traditional framework of Oracle performance tuning. However, it was not effective enough in spinlocks troubleshooting.
%Figure \ref{fig:lcont}
% illustrates the need for latch performance diagnostics and tuning. 
% During this incident the entire Oracle instance was hung due to  heavy
% \textbf{''cache buffers chains''} latch contention.

\subsection{The Tool}

%The goals of this work are:
%Explore one of Oracle serialization mechanisms: latches. Explore latch efficiency %and possibilities of diagnostics and performance tuning. Explore how to %interpret latch related performance counters. Explore latch spinning and %waiting policies.Explore influence of Oracle parameters and adjustment of %the number of spins for the latch before waiting

To discover how the Oracle latch works, we need the tool. Oracle Wait Interface allows us to explore the waits only. Oracle \textbf{X\$\slash V\$} tables instrument the latch acquisition and give us performance counters. To see how latch works through time and to observe short duration events, we need something like stroboscope in physics. Likely, such tool exists in Oracle Solaris\texttrademark. This is DTrace, Solaris 10 Dynamic Tracing framework \cite{dtrace}.

DTrace is event-driven, kernel-based instrumentation that can   see and measure all OS activity. It allows defining the \textbf{probes} (triggers) to trap and write the handlers (\textbf{actions}) using dynamically interpreted C-like language. No application changes needed to use DTrace.  This is very similar to triggers in database technologies. 

DTrace provides more than 40000 probes in Solaris kernel and ability to instrument every user instruction. It describes the triggering probe in a four-field format: \textbf{provider:module:function:name}.

A provider is a methodology for instrumenting the system: \textbf{pid, fbt, syscall, sysinfo, vminfo} \ldots

 If one need to set trigger inside the \textbf{oracle} process with Solaris \textbf{spid} 16444, to fire on entry to function \textbf{kslgetl} (get exclusive latch), the probe description will be \textbf{pid16444:oracle:kslgetl:entry}. Predicate   and action  of probe will filter, aggregate and print out the data. All the scripts used in this work are the collections of such triggers.

 Unlike standard tracing tools, DTrace works in Solaris kernel. When \textbf{oracle} process entered probe function, the execution went to Solaris kernel and the DTrace filled buffers with the data. The \textbf{dtrace} program printed out these buffers.

Kernel based tracing is more stable and have less overhead then userland. DTrace sees all the system activity and can take into account the {\em{}''unaccounted for''} userland tracing time associated with kernel calls, scheduling, etc.

DTrace allowed this work to investigate how Oracle latches perform in real time:
\begin{itemize}
\item  Count the latch spins
\item    Trace how the latch waits
\item   Measure times and distributions
\item   Compute additional latch statistics
\end{itemize}

The following next sections describe Oracle performance tuning and database administrator's related results. Reader interested in  mathematical estimations may proceed directly to section \ref{sec:cont}

\section{Oracle latch instrumentation}

It was known that the Oracle server uses \textbf{kslgetl} √- {\em{}Kernel Service Lock Management Get Latch} function to acquire the latch. DTrace reveals other latch interface routines:

\begin{itemize}
\item\textbf{kslgetl(laddr, wait, why, where)} -- get exclusive latch
\item\textbf{kslg2c(l1,l2,trc,why, where)} -- get two excl. child latches
\item\textbf{kslgetsl(laddr,wait,why,where,mode)} -- get shared latch.  In Oracle 11g -- \textbf{ksl\_get\_shared\_latch()} 
\item\textbf{kslg2cs(l1,l2,mode,trc,why, where))} -- get two shared child latches
\item\textbf{kslgpl(laddr,comment,why,where)} -- get parent and all childs
\item\textbf{kslfre(laddr)}   -- free the latch 
\end{itemize}

\normalsize

Fortunately        Oracle gave us possibility to do the same using \textbf{oradebug call} utility. {It is possible to acquire the latch manually}. This is very useful  to simulate latch related hangs and contention.
\small\begin{verbatim}
SQL>oradebug call kslgetl <laddress> <wait> <why> <where>
\end{verbatim}
\normalsize

DTrace scripts also demonstrated the meaning of arguments:
\begin{itemize}
\item  \textbf{laddres} --√ address of latch in SGA
\item  \textbf{wait} --√ flag for no-wait or wait latch acquisition
\item  \textbf{where} --√ integer code for location from where the latch is acquired.
\item  \textbf{why} ---  integer context of why the latch is acquiring at this ⌠where■. 
\item  \textbf{mode} --√ requesting state for shared lathes. 8 -√ SHARED mode. 16 √- EXCLUSIVE mode
\end{itemize}

\textbf{''Where''} and \textbf{''why''} parameters are using for the instrumentation of latch get.

  Integer \textbf{''where''} value is the reason for latch acquisition. This is the index in an array of ''locations'' strings that literally describes \textbf{''where''}. Oracle externalizes this array to SQL in  \textbf{x\$ksllw} fixed table.  These strings the database administrators are commonly see
in \textbf{v\$latch\_misses} and AWR/Statspack reports.

Fixed view v\textbf{\$latch\_misses} is based on \textbf{x\$kslwsc} fixed table. In this table Oracle maintains an array of counters for latch misses by \textbf{''where''} location. 

\textbf{''Why''} parameter is named \textbf{''Context saved from call''} in dumps. It specifies why the latch is acquired at this \textbf{''where''}.
% Its meaning depends on latch and ⌠where■. For example, ⌠why■ contains DBA %address of block that accessed under protection of cache buffers chain latch. % Tanel Poder elegantly used this to investigate the root cause of cache %buffers chain latch contention.

\textbf{''Where''} and \textbf{''why''} parameters instrument the latch get. When the latch will be acquired, Oracle saves these values into the latch structure.   Oracle 11g externalizes latch structures in \textbf{x\$kslltr\_parent} and \textbf{x\$kslltr\_children} fixed tables for parent and child latches respectively. Versions 10g and before used \textbf{x\$ksllt} table. Fixed views \textbf{v\$latch} and \textbf{v\$latch\_children} were created on these tables.

\textbf{''Where''} and \textbf{''why''} parameters for last latch acquisition may be seen in \textbf{kslltwhr} and \textbf{kslltwhy} columns of these tables. 
Fixed table \textbf{x\$ksuprlat} shows latches that processes are currently holding. View \textbf{v\$latchholder} created on it. Again, \textbf{''where''} and \textbf{''why''} parameters of latch get present  in \textbf{ksulawhr} and \textbf{ksulawhy} columns.

When Oracle process waits (sleeps) for the latch, it puts latch address into \textbf{ksllawat}, \textbf{''where''} and \textbf{''why''} values into \textbf{ksllawer} and \textbf{ksllawhy} columns of corresponding  \textbf{x\$ksupr} row. This is the fixed table behind the  \textbf{v\$process} view. These columns are extremely useful when  exploring why the processes contend for the latch.
        
\begin{figure}\label{fig:kslla}
\begin{center}
\setlength\fboxsep{0pt}
\setlength\fboxrule{0.5pt}
\fbox{
\includegraphics[width=0.8\linewidth]{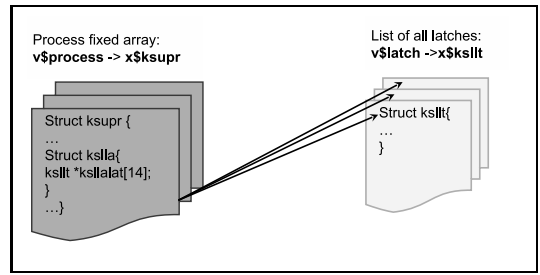}
}

\caption{Latch is holding by process, not session}
\end{center}
\end{figure}

This is illustrated on Figure \ref{fig:kslla}. 
%\vspace*{-5mm}

In summary, Oracle instruments the latch acquisition in
\textbf{x\$ksupr} fields:
\begin{itemize}
\item \textbf{ksllalaq} √-- address of latch acquiring. Populated during immediate get (and spin before 11g)
\item   \textbf{ksllawat} --- latch being waited for. 
\item    \textbf{ksllawhy} --√ ⌠why■ for the latch being waited for
\item   \textbf{ksllawere} --√ ⌠where■ for the latch being waited for
\item    \textbf{ksllalow} --√ bit array of levels of currently holding latches
\item    \textbf{ksllaspn} --- latch this process is spinning on. Not populated since 8.1
\item    \textbf{ksllaps\%} --- inter-process post statistics
\end{itemize}

\subsection{The latch structure -√ \textbf{ksllt}}

Latch structure is named \textbf{ksllt} in Oracle fixed tables. It contains the latch location itself, \textbf{''where''} and \textbf{''why''} values, latch level, latch number, class, statistics, wait list header and other attributes. 

\hfill Table \ref{tab:ksllt}. {Latch size by Oracle version}
\vspace*{-5mm}
\small{
\begin{center}\label{tab:ksllt}
\begin{tabular}{|l|c|c|c|}
\hline
&&&\\
Version&Unix 32bit&Unix 64bit&Windows 32bit\\
\hline 7.3.4&92&--&120\\
\hline 8.0.6&104&--&104\\
\hline 8.1.7&104&144&104\\
\hline 9.0.1&?&200&160\\
\hline 9.2.0&196&240&200\\
\hline 10.1.0&?&256&208\\
\hline 10.2.0-11.2.0.2&100&160&104\\
\hline
\end{tabular}
\end{center}
}
\normalsize

Contrary to popular believe Oracle latches were significantly evolved through the last decade. Not only additional statistics appeared (and disappeared) and new (shared)  latch type was introduced,  the latch  itself was changed.
Table \ref{tab:ksllt} shows how the latch structure size changed  by Oracle version.
The \textbf{ksllt} size decreased in 10.2 because Oracle made obsolete many latch statistics. 

Oracle latch is not just a single memory location. Before Oracle 11g the value of first latch byte (word for shared latches) was used to determine latch state:
\begin{itemize}
\item\textbf{0x00} --√ latch is free.
\item\textbf{0xFF} √-- exclusive latch is busy. Was 0x01 in Oracle 7.
\item\textbf{0x01,0x02,etc.} --- shared latch holding by 1,2, etc. processes simultaneously.
\item\textbf{0x20000000 $|$ pid} --- shared latch holding exclusively.
\end{itemize}
In Oracle 11g the first exclusive latch word represents the Oracle \textbf{pid} of the latch holder:
\begin{itemize}
\item\textbf{0x00} --√ latch free.
\item\textbf{0x12} --√ Oracle process with \textbf{pid} 18 holds the exclusive latch.
\end{itemize}

%%%

\subsection{Latch attributes}
According to Oracle Documentation and DTrace traces, each latch has at least the following flags and attributes:
\begin{itemize}
\item   \textbf{Name} --- Latch name as appeared in \textbf{V\$} views
\item   \textbf{SHR} --- Is the latch {\em{}Shared}? Shared latch is ⌠Read-Write■ spinlock.
\item \textbf{PAR} --- Is the latch {\em{}Solitary} or {\em{}Parent} for the family of child latches? Both parent and child latches share the same latch name. The parent latch can be gotten independently, but may act as a master latch when acquired in special mode  in \textbf{kslgpl()}. 
\item    \textbf{G2C} --- Can two child latches be simultaneously requested in wait mode?
\item   \textbf{LNG} --- Is wait posting used for this latch? Obsolete since Oracle 9.2.
\item  \textbf{UFS} --- Is the latch Ultrafast? It will not increment miss statistics when STATISTICS\_LEVEL=BASIC.  10.2 and above
\item    \textbf{Level}. 0-14. To prevent deadlocks latches can be requested  \emph{only in increasing level order}.
\item     \textbf{Class}. 0-7. Spin and wait class assigned to the latch. Oracle 9.2 and above.
\end{itemize}
Evolution of Oracle latches   is summarized in table \ref{tab:kslld}. 

\hfill Table \ref{tab:kslld}. {Latch attributes by Oracle version}

\vspace*{-5mm}
\small{
\begin{center}\label{tab:kslld}
\begin{tabular}{|l|c|c|c|c|c|c|}
\hline 
&&&&&&\\
Oracle &Number of&       PAR &    G2C&     LNG  &   UFS   &  SHR\\
version&latches&&&&&\\
\hline 7.3.4.0  &53   &14&2&3&---&---\\
\hline 8.0.6.3  & 80  &    21    &  7      & 3    &   ---       &3\\
\hline 8.1.7.4  &152  &    48    &  19     &4     &  ---       &9\\
\hline 9.2.0.8  &242  &    79    &  37     &---   &    ---     &  19\\
\hline 10.2.0.2 &385  &   114    & 55      &---   &    4       &47\\
\hline 10.2.0.3 &388  &   117    & 58      &---   &    4       &48\\
\hline 10.2.0.4 &394  &   117    & 59      &---   &    4       &50\\
\hline 11.1.0.6 &496  &   145    & 67      &---   &    6       &81\\
\hline 11.1.0.7 &502  &   145    & 67      &---   &    6       &83\\
\hline 11.2.0.1 &535  &   149    & 70      &---   &    6       &86\\
\hline
\end{tabular}
\end{center}
}\normalsize

To prevent deadlocks  Oracle process can acquire latches only with level higher than it currently holding.  At the same level, the process can request the second \textbf{G2C} latch child X in wait mode after obtaining child Y, {\em{}if and only if the child number of X $<$ child number of Y.} If these rules are broken, the Oracle process raises ORA-600 errors.

''Rising level'' rule leads to ''trees'' of processes waiting for and holding the latches. Due to this rule the contention for higher level latches frequently exacerbates contention for lower level latches. These trees can be seen  by direct SGA access programs.

Each latch can be assigned to one of 8 classes  with different spin and wait policies. By default, all the latches belong to class 0.
The only exception is  \textbf{''process allocation latch''}, which belongs to class 2.
Latch assignment to classes is controlled by initialization parameter  \textbf{\_LATCH\_CLASSES}. Latch class spinning and waiting policies can be adjusted by 8 parameters named \textbf{\_LATCH\_CLASS\_0} to \textbf{\_LATCH\_CLASS\_7}.

\subsection{Latch Acquisition in Wait Mode}

According to contemporary Oracle 11.2 Documentation, latch wait get (kslgetl(laddress,1,\ldots))  proceeds through the following phases: 
\begin{itemize}
\item  One fast \textbf{Immediate get}, no spin. 
\item  \textbf{Spin get}: check the latch upto \textbf{\_SPIN\_COUNT} times. 
\item  \textbf{Sleep} on \textbf{''latch free''} wait event with exponential backoff.
\item   Repeat.
\end{itemize}
It occurs that such algorithm was really used ten years ago in Oracle versions 7.3-8.1. For example, look at Oracle 8i latch get code flow using Dtrace:
\small\begin{verbatim}
kslgetl(0x200058F8,1,2,3)   -KSL GET exclusive Latch 
 kslges(0x200058F8, ...)    -wait get 
  skgsltst(0x200058F8)   ...   call repeated 2000 times
  pollsys(...,timeout=10 ms)- Sleep 1
   skgsltst(0x200058F8)  ...   call repeated 2000 times
  pollsys(...,timeout=10 ms)- Sleep 2
   skgsltst(0x200058F8)  ...   call repeated 2000 times
  pollsys(...,timeout=10 ms)- Sleep 3
   skgsltst(0x200058F8)  ...   call repeated 2000 times
  pollsys(...,timeout=30 ms)- Sleep 4 ...
\end{verbatim}
\normalsize
The 2000 cycles is the value of \textbf{SPIN\_COUNT} initialization parameter. This value could be changed dynamically without Oracle instance restart.

Corresponding Oracle event 10046 trace \cite{millsap}  is:
\small\begin{verbatim}
WAIT #0: nam='latch free' ela=1 p1=536893688 p2=29 p3=0
WAIT #0: nam='latch free' ela=1 p1=536893688 p2=29 p3=1
WAIT #0: nam='latch free' ela=1 p1=536893688 p2=29 p3=2
WAIT #0: nam='latch free' ela=3 p1=536893688 p2=29 p3=2
\end{verbatim}\normalsize
The sleeps timeouts demonstrate the exponential backoff:
\small\begin{verbatim}
0.01-0.01-0.01-0.03-0.03-0.07-0.07-0.15-0.23-0.39-0.39- 
     0.71-0.71-1.35-1.35-2.0-2.0-2.0-2.0...sec
\end{verbatim}\normalsize
This sequence can be almost perfectly fitted by the following formula.
\begin{equation}
\textbf{timeout}=2^{\left[(N_{wait}+1)/2\right]}-1
\end{equation}
However, such sleep for predefined time was not efficient. Typical latch holding time is less then {\em{}10 microseconds}. {\em{}Ten milliseconds}  sleep  was too large.   Most waits were for nothing, because latch already was free. In addition,  repeating sleeps resulted in many unnecessary spins, burned CPU and provokes CPU thrashing.

It was not surprising that in Oracle 9.2-11g exclusive latch get was changed significantly. DTrace demonstrates its code flow:
\small\begin{verbatim}
kslgetl(0x50006318, 1)
 sskgslgf(0x50006318)= 0     -Immediate latch get
 kslges(0x50006318, ...)     -Wait latch get
  skgslsgts(...,0x50006318)  -Spin latch get
   sskgslspin(0x50006318)... - repeated 20000 cycles
    kskthbwt(0x0)
    kslwlmod()               - set up Wait List
    sskgslgf(0x50006318)= 0  -Immediate latch get
    skgpwwait                -Sleep latch get
     semop(11, {17,-1,0}, 1)
\end{verbatim}\normalsize
Note the \textbf{semop()} operating system call. This is infinite wait until posted. This operating system call will block the process until another process posts it during latch release.

Therefore, in Oracle 9.2-11.2, all the latches in default class 0 rely on wait posting. Latch is sleeping without any timeout. This is more efficient than previous algorithm. Contemporary latch statistics shows that most latch waits is less then 1 ms now. In addition, spinning once reduce CPU consumption. 

However, this introduces a problem. If wakeup post is lost in OS, waiters will sleep infinitely. This was common problem in earlier 2.6.9 Linux kernels. Such losses can lead to instance hang because the process will never be woken up. Oracle solves this problem by \textbf{\_ENABLE\_RELIABLE\_LATCH\_WAITS} parameter. It changes the \textbf{semop()} system call to \textbf{semtimedop()} call with 0.3 sec timeout. 

Latches assigned to non-default class wait until timeout. Number of spins and duration of sleeps for class \textbf{X} are determined by corresponding \textbf{\_LATCH\_CLASS\_X} parameter, which is a string of:
\begin{center}
 \textbf{''Spin Yield Waittime Sleep0 Sleep1 \ldots Sleep7''}
\end{center}
%Non
%\begin{verbatim} 
%        If ''Yield'' !=0  repeat ''Yield'' times:
%          Loop up to ''Spins'' cycles 
%          Yield CPU using yield() 
%        Sleep for ''SleepX'' usecs
%        Then spin again  ┘
%\end{verbatim}
Detailed description of non-default latch classes can be found in \cite{asn}.

DTrace demonstrated that by default the process spins for exclusive latch for 20000 cycles. This is determined by static \textbf{\_LATCH\_CLASS\_0} initialization parameter. The \textbf{\_SPIN\_COUNT} parameter (by default 2000) is effectively static for exclusive latches \cite{asn}. Therefore spin count for {\em{}exclusive latches} can not be changed without instance restart.

Further DTrace investigations showed that {\em{}shared latch} spin in Oracle 9.2-11g is governed by \textbf{\_SPIN\_COUNT} value and can be dynamically tuned. Experiments demonstrated that
       \textbf{X} mode shared latch get spins by default up to 4000 cycles.
       \textbf{S} mode does not spin at all (or spins in unknown way).  Discussion how Oracle shared latch works can be found in \cite{asn}.  The results are summarized in table \ref{tab:sharedl}.

\hfill Table \ref{tab:sharedl}. {Shared latch acquisition}

\vspace*{-5mm}
\small{
\begin{center}\label{tab:sharedl}
\begin{tabular}{|l|c|l|}
\hline
&\textbf{S mode get} &\textbf{X mode get}\\
\hline \textbf{Held in S mode}&Compatible&2*\_SPIN\_COUNT\\
\hline \textbf{Held in X mode}&0&2*\_SPIN\_COUNT\\
\hline \textbf{Blocking mode}&0&2*\_SPIN\_COUNT\\
\hline
\end{tabular}
\end{center}
}\normalsize

\subsubsection{Latch Release} 

 Oracle process releases the latch  in \textbf{kslfre(laddr)}. To deal with invalidation storms \cite{Anderson90}, the process releases the latch nonatomically.
 Then it sets up {\em{}memory barrier} using atomic operation on address individual to each process.   This requires less bus invalidation and ensures propagation of latch release to other local caches. 

  This is not fair policy. Latch spinners on the local CPU board have the preference. However, this is more efficient then atomic release. Finally the  process posts  first process in the list of waiters. 
 
\section{The latch contention}\label{sec:cont}
\subsection{Raw latch statistic counters}
%Now we can use the knowledge how the Oracle process spins and waits to   analyse latch statistics. 
 
 Latch statistics is the tool to estimate whether  the latch acquisition works efficiently or we need to tune it. Oracle counts  a broad range of latch related statistics. Table \ref{tab:lstat} contains description of \textbf{v\$latch} statistics columns from contemporary Oracle documentation \cite{Concepts112}.

Oracle collects more statistics then are usually consumed by classic queuing models. 

\hfill Table \ref{tab:lstat}. {Latch statistics}

%\vspace*{-5mm}
\begin{center}\label{tab:lstat}
\small{
\begin{tabular}{|p{0.15\linewidth}|p{0.4\linewidth}|p{0.3\linewidth}|}
\hline
&&\\
\textbf{Statistic:}&\textbf{Documentation description:}&\textbf{When and how it is changed:}\\
&&\\
\hline 
GETS&Number of times the latch was requested in willing-to-wait mode&Incremented by one after latch acquisition \\
\hline
MISSES&    Number of times the latch was requested in willing-to-wait mode and the requestor had to wait&      Incremented by one after latch acquisition if miss occurred\\
\hline
SLEEPS&   Number of times a willing-to-wait latch request resulted in a session sleeping while waiting for the latch&       Incremented by number of times process slept during latch acquisition\\
\hline
SPIN\-\_GETS
&       Willing-to-wait latch requests, which missed the first try but succeeded while spinning&      Incremented by one after latch acquisition if miss but not sleep occured. Counts only the first spin\\
\hline
WAIT\-\_TIME
&     Elapsed time spent waiting for the latch (in microseconds)&         Incremented by wait time spent during latch acquisition.\\
\hline
IMMED\-IATE\-\_GETS
&   Number of times a latch was requested in no-wait mode&   Incremented by one after each no-wait latch get\\
\hline
IMMED\-IATE\-\_MISSES
&  Number of times a no-wait latch request did not succeed&         Incremented by one after unsuccessful no-wait latch get\\
\hline
\end{tabular}
}
\end{center}
\normalsize

Since version 10.2 many previously collected latch statistics have been deprecated. We have lost important additional information about latch performance. Here I will discuss the remaining statistics set.

As was demonstrated in previous chapter, since version 9.2 Oracle uses completely new latch acquisition algorithm:
\begin{verbatim}
Immediate latch get
 Spin latch get
  Add the process to waiters queue  
   Sleep until posted
\end{verbatim}

%We should analyze latch statistics from this new algorithm point of view.

GETS, MISSES, etc. are the integral statistics counted from the startup of the instance. These values depend on complete workload history. AWR and Statspack reports show changes of integral statistics per snapshot interval. Usually these values are ''averaged by hour'', which is much longer then typical latch performance spike.

Another problem with AWR\slash Statspack report is averaging over child latches. By default AWR gathers only summary data from \textbf{v\$latch}. This greatly distorts latch efficiency coefficients. {\em{}The latch statistics should not be averaged over child latches.}

 To avoid averaging distortions the following analysis uses the latch statistics from \textbf{v\$latch\_parent} and \textbf{v\$latch\_children} (or \textbf{x\$ksllt} in Oracle version less then 11g)

The current workload is characterized by differential latch statistics and ratios. 

\hfill Table 3.2 {Differential (point in time) latch statistics}
\vspace{-3mm}
\begin{center}\label{tab:dstat}
\small{
\begin{tabular}{|p{0.22\linewidth}|l|l|}
\hline
&&\\
\textbf{Description:}&\textbf{Definition:}&\textbf{AWR equivalent:}\\
&&\\
\hline%\vbox{\vspace{10pt}}
&&\\
%Latch requests 
\textbf{Arrival rate}&
%\vbox{\vspace{15pt}}
$%\displaystyle
\ \lambda=\frac{\Delta GETS}{\Delta time}$&$\ \frac{''Get\ Requests''}{''Snap\ Time\ (Elapsed)''}$\\
&&\\
\hline
&&\\ \textbf{Gets efficiency}&$\ \rho=\frac{\Delta MISSES}{\Delta GETS}$&$\ ''Pct\ Get\ Miss''/100$\\
%&&\\
\hline
&&\\ \textbf{Sleeps ratio}&$\ \kappa=\frac{\Delta SLEEPS}{\Delta MISSES}$&$\ ''Avg\ Slps\ /Miss''$\\
&&\\
\hline
&&\\ \textbf{Wait time per second}&$\ W =\frac{\Delta WAIT\_TIME}{10^6\Delta time}%*10^{-6}
$&$\ \frac{''Wait\ Time\ (s)''}{''Snap\ Time\ (Elapsed)''}$\\
%&&\\
\hline
&&\\
 \textbf{Spin\newline efficiency}&$\ \sigma=\frac{\Delta SPIN\_GETS}{\Delta MISSES}$&$\ \frac{''Spin\ Gets''}{''Misses''}$\\
%&&\\
\hline\end{tabular}
}
\end{center}
\normalsize

%%%%

There exist several ways to choose the basic set of differential statistics. I will use the most close to AWR\slash Statspack way containing ''Arrival rate'', ''Gets efficiency'', ''Spin efficiency'', ''Sleeps ratio'' and ''Wait time per second''. Table 3.2 defines  these quantities.

This work analyzes  only {\em{}wait} latch gets. The no-wait {\em{}(IMMEDIATE\_\ldots)} gets add some complexity only for several latches. I will also assume $ {\Delta time}$ to be small enough that workload do not change significantly.

Other statistics reported by AWR depend on these key statistics:
\begin{itemize}
\item Latch \textbf{miss rate} is $ \frac{\Delta MISSES}{\Delta time}=\rho\lambda$. %This is ''Misses'' from AWR divided by snapshot interval
\item Latch \textbf{waits (sleeps) rate} is $ \frac{\Delta SLEEPS}{\Delta time}=\kappa\rho\lambda$.
% This is per second value of ''Sleeps'' from AWR
\end{itemize}
From the queuing theory point of view, the latch is G/G/1/(SIRO+FIFO) system with interesting queue discipline including {\em{}Serve In Random Order} spin and {\em{}First In First Out} sleep. Using the latch statistics, I can roughly estimate queuing characteristics of latch. I expect that the accuracy of such estimations is about 20-30\%. 

As a first approximation, I will assume that incoming latch requests stream is Poisson  and latch holding (service) times are exponentially distributed. Therefore, our first latch model will be M/M/1/(SIRO+FIFO). 

Oracle measures more statistics  then usually consumed by classic queuing models.  It is interesting what these additional statistics can be used for.

\subsection{Average service time:}

The PASTA (Poisson Arrivals See Time Averages) \cite{kleinrock} property  connects $ \rho$ ratio with the latch utilization. For Poisson streams the latch gets efficiency should be equal to utilization:
\begin{equation} 
\rho=\frac{\Delta misses}{\Delta gets}\approx  U=\frac{\Delta latch\ hold\ time}{\Delta time}
\end{equation}
However, this is not exact for server with finite number of processors. The Oracle process occupies the CPU while acquiring the latch. As a result, the latch get see the utilization induced by other $ N_{CPU}-1$ processors only. Compare this with  MVA \cite{MVA} arrival theorem. In some benchmarks there may be only $ N_{proc}\leq N_{CPU}$ Oracle shadow processes that generates the latch load. In such case we should substitute $ N_{proc}$ instead $ N_{CPU}$ in the following estimate:

\begin{equation}
 \rho \simeq \left( 1-\frac{1}{min(N_{CPU},N_{proc})}\right) U=\frac{1}{\eta}U
 %;\quad\eta=\frac{min(N_{CPU},N_{proc})}{min(N_{CPU},N_{proc})-1}
\end{equation}
Here I introduced the  the 
\[ 
 \eta=\frac{min(N_{CPU},N_{proc})}{min(N_{CPU},N_{proc})-1}
\]
 multiplier to correct naive utilization estimation. Clearly, the $ \eta$ multiplier confirms that the entire approach is inapplicable to single CPU machine. Really $ \eta$ significantly differs from one only during demonstrations on my Dual-Core notebook. For servers its impact is below precision of my estimates.  For example for small 8 CPU server the $ \eta$ multiplier  adds only 14\% correction. 

We can experimentally check the accuracy of these formulas and, therefore, Poisson arrivals approximation. $U$ can be independently measured by sampling of \textbf{v\$latchholder}. The \textbf{latchprofx.sql}   script by Tanel Poder \cite{lp} did this at high frequency. Within our accuracy we can expect that $ \rho$ and $U$ should be at least of the same order. 

%To calculate the complete statistics set, I wrote  \textbf{latch\_stats\_10g.sql} %script \cite{asn}. The script uses latch address as the parameter. It computes %differential latch statistics for small interval (30sec) and estimates latch %utilization and queue length by sampling. Note that this is only demonstration %script. To measure such quantities accurately  we need to perform statistical %analysis and least squares fits.
 
We know that $ U= \lambda S$, where $ S$ is average service (latch holding) time. This allows us to estimate the latch holding  time as:
\begin{equation}
 S=\frac{\eta\rho}{\lambda}
\end{equation}

This is interesting. We obtained the first estimation of latch holding time directly from statistics. In AWR terms this formula looks like 
\begin{displaymath}
 S=\eta\frac{''Pct\ Get\ Miss''\ \times\ ''Snap\ Time''}{100\ \times\  ''Get\ Requests''}
%=\eta\frac{''Misses''\times''Snap\ Time''}{''Gets''^2}
\end{displaymath} 

%The inverse of service time is the service rate $ \mu = 1/S=\frac{\lambda}{\eta\rho}$. %This is also the estimation of the upper limit for latch request frequency %the instance can serve. 

\subsection{ Wait time:}

Look more closely on the summary wait time per second $ W$.  Each latch acquisition increments the \textbf{WAIT\_TIME} statistics by amount of time it waited for the latch. According to the Little law, average latch sleeping time is related the length of wait (sleep) queue: 
\begin{displaymath}
L = \lambda_{waits}\times  \left\langle average\ wait\ time \right\rangle = \lambda \rho \kappa \times \delta (Wait\_Time) 
\end{displaymath}

The right hand side of this identity is exactly the  \textbf{''wait time per second''} statistic. Therefore, actually:
\begin{equation} 
W\equiv L
\end{equation}

We can experimentally confirm this conclusion because $ L$ can be independently measured by sampling of \textbf{v\$process.latchwait} column. 

\subsection{Recurrent sleeps:}

In ideal situation, the process spins and sleeps only once. Consequently, the latch statistics should satisfy the following identity: 
\begin{equation} MISSES\ =\ SPIN\_GETS\ +\  SLEEPS 
\end{equation}
 Or, equivalently:
\begin{equation} 1=\sigma+\kappa
\end{equation} 
In reality, some processes had to sleep for the latch several times. This occurred when the sleeping process was posted, but  another process got the latch before the first process received the CPU. The awakened process spins and sleeps again. As a results the previously equality became invalid.

Before version 10.2 Oracle directly counted these sequential waits in separate \textbf{SLEEP1-SLEEP3} statistics. Since 10.2 these statistics became obsolete. However, we can estimate the rate of such ''sleep misses'' from other basic statistics. The recurrent sleep increments only the \textbf{SLEEPS} counter. The \textbf{SPIN\_GETS} statistics not changed. The $ \sigma+\kappa-1$ is the ratio of inefficient latch sleeps to misses. The ratio of ''unsuccessful sleep'' to ''sleeps'' is given by:
\begin{equation}
\rm{Recurrent\ sleeps\ ratio} =\frac{\sigma+\kappa-1}{\kappa}
\end{equation}
Normally this ratio should be close to $\rho$. Frequent ''unsuccessful sleeps'' are inefficient and may be a  symptom of OS waits posting problems or bursty workload.

\subsection{Latch acquisition time:}

Average latch acquisition time is the sum of {\em{}spin} time and {\em{}wait} time. Oracle does not directly measure the spin time.  However, we can measure it on Solaris platform using DTrace.

On other platforms, we should rely on statistics. Fortunately in Oracle 9.2-10.2 one can count the average number of spinning processes by sampling \textbf{x\$ksupr.ksllalaq}. The process set this column equal to address of acquired latch during active phase of latch get. Oracle 8i and before even fill the \textbf{v\$process.latchspin} during latch spinning. 

Little law allows us to connect average number of spinning processes with the spinning time:
\begin{equation} N_s = \lambda T_s
\end{equation}
As a result the average latch acquisition time is:
\begin{equation} T_a= \lambda^{-1} (N_s+W)
\end{equation}
Note that according to general queuing theory the  ''Serve In Random Order'' discipline of latch spin does not affect average latch acquisition time. It is independent on queuing discipline. In steady state, the number of processes served during the passage of incoming request through the system should be equal to the number of spinning and waiting processes.

In Oracle 11g the latch spin is no longer instrumented due to a bug. The 11g spin is invisible for SQL. This  do not allow us to estimate $ N_s$ and related quantities. 

\subsection{Comparison of results}

Let me compare the results of DTrace measurements and latch statistics.
%Unique Solaris DTrace allows us directly measure latch acquisition and holding %times and their distributions. 
%For the following demonstration, I artifically induced the library cache
%contention repeatedly selecting from \textbf{v\$sql\_plan}. 
Typical demonstration results for our 2 CPU  X86  server are:
\small\begin{verbatim}
/usr/sbin/dtrace -s latch_times.d -p 17242 0x5B7C75F8
...
latch gets traced: 165180 
''Library cache latch'', address=5b7c75f8
Acquisition time:
value  ------------- Distribution ------------- count
    4096 |                                         0
    8192 |@@                                       7324
   16384 |@@@@@@@@@@@@@@@@@@@@@@@@@@@@@@@@@@@@     151748
   32768 |@                                        4493
   65536 |                                         1676
  131072 |                                         988
  262144 |                                         464
  524288 |                                         225
 1048576 |                                         211
 2097152 |                                         53
 4194304 |                                         21
 8388608 |                                         1
16777216 |                                         1
33554432 |                                         0

Holding time:
value  ------------- Distribution ------------- count
    8192 |                                         0
   16384 |@@@@@@@@@@@@@@@@@@@@@@@@@                105976
   32768 |@@@@@@@@@@@@                             50877
   65536 |@@                                       6962
  131072 |                                         1986
  262144 |                                         829
  524288 |                                         330
 1048576 |                                         205
 2097152 |                                         34
 4194304 |                                         6
 8388608 |                                         0

Average acquisition time =26 us
Average holding time =37 us
\end{verbatim}\normalsize
 The above histograms show latch acquisition and holding time distributions in logarithmic scale.  Values are in nanoseconds. Compare the above average times with the results of latch statistics analysis under the same conditions:
\small\begin{verbatim}
Latch statistics  for  0x5B7C75F8   
''library cache''  level#=5   child#=1
Requests rate:       lambda=  20812.2 Hz
Miss /get:              rho=  0.078
Est. Utilization:   eta*rho=  0.156
Sampled   Utilization:    U=  0.143
Slps /Miss:      kappa=    0.013
Wait_time/sec:       W=    0.025
Sampled queue length L=    0.043
Spin_gets/miss:  sigma=    0.987
Sampled spinnning:  Ns=    0.123
Derived statistics:
Secondary sleeps ratio =  0.01
Avg latch holding time =      7.5 us
         sleeping time =      1.2 us
      acquisition time =      7.2 us
\end{verbatim}\normalsize
We can see that $ \eta\rho$ and $ W$ are close to sampled $ U$ and $ L$ respectively. The holding and acquisition times from both methods are of the same order. Since both methods are intrusive, this is remarkable agreement. Measurements of latch times and distributions for demo and production workloads conclude that:

%On the other hand for the purpose of contention diagnostics the main result
%of this post is:

{\em{}The latch holding time for the contemporary servers should be normally in microseconds range.}

\section{Latch contention  in Oracle 9.2-11g}

 Latch contention should be suspected if the latch wait events are observed in ⌠Top 5 Timed Events■ AWR section. Look for the latches with highest $W$.
Symptoms of contention for the latch are highly variable. Most commonly observed include:
\begin{itemize}
\item $W > 0.1$ sec/sec
\item Utilization     $>$ 10\%
\item Acquisition (or sleeping) time significantly greater then holding time
\end{itemize}

 \textbf{V\$latch\_misses} fixed view and \textbf{latchprofx.sql} script  by Tanel Poder \cite{lp} reveal {\em{}''where''} the contention arise.
One should always take into account that contention for a high-level latch frequently exacerbates contention for lower-level latches \cite{adams}.

How treat the latch contention? During the last 15 years, the latch performance tuning was focused  on application tuning and reducing the latch demand. To achieve this one need to tune the SQL operators, use bind variables, change the physical schema, etc\ldots Classic Oracle Performance books explore these  topics \cite{adams,millsap,OWI}.

However, this tuning methodology may be too expensive and even require complete application rewrite. This work explores complementary possibility of changing  \textbf{\_SPIN\_COUNT}. This commonly treated as old style tuning, which should be avoided at any means. Increasing of spin count may leed to waste of CPU. However, nowadays the CPU power is cheap. We may already have enough free  resources. We need to find conditions when the spin count tuning may be beneficial.

Processes spin for exclusive latch spin upto 20000 cycles, for shared latch upto 4000 cycles and infinitely for mutex. Tuning may find more optimal values for your application.

Oracle does not explicitly forbid spin count tuning. However, change of undocumented parameter should be discussed with Support.

\subsection{Spin count adjustment}

Spin count tuning depends on latch type. For {\em{}shared}  latches:
\begin{itemize}
\item Spin count can be adjusted dynamically by \textbf{\_SPIN\_COUNT} parameter. \item Good starting point is the multiple of default 2000 value.
\item Setting \textbf{\_SPIN\_COUNT} parameter in initialization file, should be accompanied by \textbf{\_LATCH\_CLASS\_0}=''20000''. Otherwise spin for exclusive latches will be greatly affected by next instance restart.
\end{itemize}
On the other hand if contention is for {\em{}exclusive} latches then:
\begin{itemize}
\item   Spin count adjustment by \textbf{\_LATCH\_CLASS\_0} parameter needs the instance restart. 
\item Good starting point is the multiple of default 20000 value.
\item  It may be preferable to increase the number of \textbf{''yields''} for class 0 latches.
\end{itemize} 

In order to tune spin count efficiently the root cause of latch contention must be diagnosed. Obviously spin count tuning will only be effective if the latch holding time $S$ is in its normal microseconds range.
At any time the number of spinning processes should remain less then the number of CPUs. 

%Elapsed time to acquire the latch will decrease only if latch \emph{''holding %time''} is less then OS \emph{''context switch time''}. 

It is a common myth that CPU consumption will raise infinitely while we increase the spin count. However, actually the process will spin up to {\em{}''residual latch holding time''}.  The next chapter will explore this.

%%%%

\section{Latch spin CPU time}

 The  spin probes the latch holding time distribution. 
To predict effect of \textbf{\_SPIN\_COUNT} tuning, let me introduce the mathematical model. It extends the model used in \cite{agarwal91}%,karlin90,karlin91}
  for general latch holding time distribution.  As a cost function, I will estimate the  CPU time consumed while spinning.

%observable   inside Oracle database quantities.

Consider a general stream of latch acquisition events. Latch was acquired by some process at time $T_k$ and released at $T_k + h_k$, $k\in \mathcal{N}$ Here $h_k$ is the latch holding time distributed with  p.d.f. $p(t)$. I will assume that both $T_k$ and $h_k$ are generally independent for any $k$ and form a recurrent stream. Furthermore, I assume here the existence of at least second moments for all the distributions.

If $ T_{k+1} < T_k+h_k$ then the latch will be busy when the next process tries to acquire it. The {\em{}latch miss} will occur.  In this case the process will spin for the latch up to time $\Delta$. The {\em{}spin get} will succeed if:
\begin{displaymath}
T_{k+1}+\Delta > T_k + h_k
\end{displaymath}

%\begin{displaymath} \left\{
%\begin{array}{l l}
%T_k+\tau_k > T_{k+1}\\
%T_{k+1}+\Delta > T_k + \tau_k\\
%\end{array} \right.
%\end{displaymath}
The process will sleep for the latch if $T_{k+1}+\Delta < T_k + h_k$.

Therefore, the conditions for latch wait acquisition phases are:
\begin{equation} \left.
\begin{array}{l l}
\text{latch miss:} &\quad T_{k+1}< T_k+h_k, \\
\text{latch spin get:} &\quad T_k + h_k -\Delta < T_{k+1} < T_k + h_k,\\
\text{latch sleep:} &\quad T_{k+1}+\Delta < T_k + h_k.\\
\end{array} \right.
\end{equation}

If the latch miss occur, then second process will observe that latch remain busy for: 
\begin{equation}
\tau_{k+1} = T_k+h_k - T_{k+1} 
\end{equation}
This is {\em{}''residual time''} \cite{kleinrock} or {\em{} time until first event} \cite{Zernike} of latch release . Its distribution differ from that of $h_k$. To reflect  this, I will add the subscript $l$ to all residual  distributions. In addition, I will omit subscript $k$ for the stationary state.

Let me denote the  probability that missed process see latch release at time less then $t$ as:
\begin{displaymath}
 P_l(\tau<t)=P_l(t)
\end{displaymath}
and probability of not releasing the latch during time $t$ is
\begin{math}
 Q_l(\tau \ge t) = 1 - P_l(\tau<t)
\end{math} 
. Therefore, the probability to spin for the latch during time less then $ t$ is
\begin{equation}
 P_{sg}(t_s < t) = \left\{
\begin{array}{l l}
P_l(\tau_k < t)&\text{when}\quad t <\Delta\\
1&\text{when}\quad t \ge \Delta\\
\end{array}
\right.
\end{equation}
and has a discontinuity in $t= \Delta$ because the process acquiring latch  never  spins more than $ \Delta$. The magnitude of this discontinuity is  $1-P_l(\Delta)$. This is  the probability of latch sleep.\hfill

\begin{figure}%{0.2\linewidth}
\begin{center}
\includegraphics[scale=0.45]{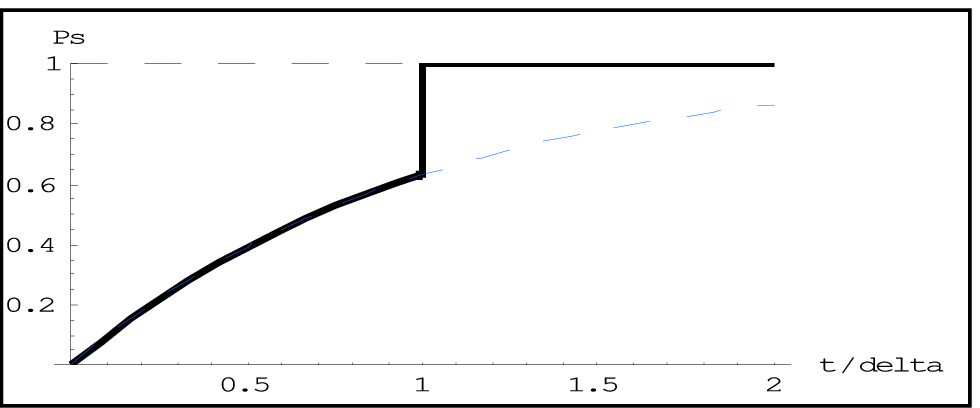}

\end{center}
\end{figure}

Therefore, the spinning probability distribution function has a bump in $\Delta$
\begin{equation} p_{sg}=p_l(t) H(\Delta-t) + (1-P_l(\Delta)) \delta(t-\Delta) \end{equation} 
Here $ H(x)$ and $ \delta(x)$ is Heaviside step and bump functions correspondingly.  
Spin efficiency is the probability to obtain latch during the spin get :
\begin{equation} \sigma= \int\limits_0^{\Delta-0} p_{sg}(t) \; \mathrm{d}t = P_l(\Delta) = 1 - Q_l(\Delta)
\end{equation}

Oracle allows measuring the average number of spinning processes. This quantity is proportional to the average CPU time spending while spinning for the latch:
\begin{equation} \Gamma_{sg}=\int\limits_0^\infty t p_{sg}(t) \; \mathrm{d}t = \int\limits_0^\Delta t p_{l}(t) \; \mathrm{d}t + \Delta (1-P_l(\Delta)) \end{equation}
Integrating by parts 
%we see:
%\begin{equation}  \Gamma_{sg}=\Delta - \int\limits_0^\Delta P_l(t) \; \mathrm{d}t %= \int\limits_0^\Delta Q_l(t) \; \mathrm{d}t 
%\end{equation}
%Another form of the same identity is:
%\begin{equation} \Gamma_{sg}=  \langle t_l \rangle - \int\limits_\Delta^\infty %Q_l \; \mathrm{d}t 
%\end{equation}
%And finally,
 both expressions may be rewritten in form:
\begin{equation} 
 \left\{\begin{array}{l}
\sigma= 1 - Q_l(\Delta)\\
\Gamma_{sg}= \Delta - \int\limits_0^\Delta P_l(t) \; \mathrm{d}t
=  \int\limits^\Delta_0 Q_l(t) \; \mathrm{d}t \\
\end{array}
\right.
\end{equation}
 or, equivalently:
\begin{equation} 
\left\{\begin{array}{l}
\sigma= 1 - Q_l(\Delta)\\
\Gamma_{sg}=  \langle t_l \rangle - \int\limits_\Delta^\infty Q_l \; \mathrm{d}t \\
\end{array}
\right.
\end{equation}

According to classic considerations from the renewal theory \cite{kleinrock}, the distribution of residual time %{\em{}until the latch release}
 is the transformed latch holding time distribution:
\begin{wrapfigure}{r}{0.27\textwidth}
\begin{center}
\vskip8mm\includegraphics[scale=0.35]{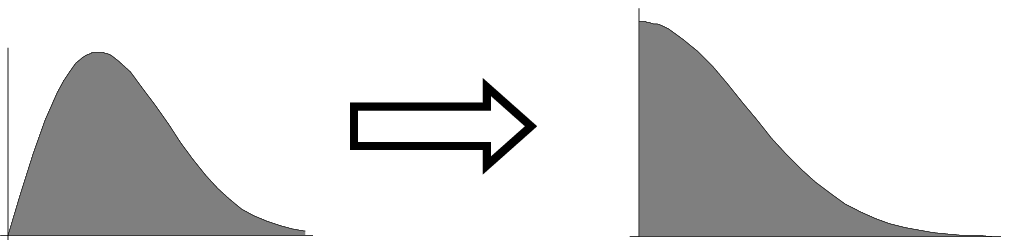}
\end{center}
\end{wrapfigure}

\begin{displaymath}
 p_l(t) = \frac{1}{\langle t \rangle} (1- P(t)) 
\end{displaymath}
The average residual latch holding time is %well-known:
\begin{math} 
 \langle t_l \rangle =\frac{\langle t^2 \rangle}{\langle 2 t \rangle }
\end{math}.
Incorporating this into previous formulas for spin efficiency and CPU time results in:
%gives:
\begin{equation} 
 \left\{\begin{array}{l}
\sigma= \frac{1}{\langle t \rangle}\int\limits^\Delta_0 Q(t) \mathrm{d}t \\
\Gamma_{sg}=  \frac{1}{\langle t \rangle}\int\limits^\Delta_0 \mathrm{d}t \int\limits^\infty_t Q(z) \; \mathrm{d}z \\
\end{array}
\right.
\end{equation}
These nice formulas encourage us that observables explored are not artifacts:
\begin{equation} \label{eq:triplei}
 \left\{\begin{array}{l}
\sigma= \frac{1}{\langle t \rangle}\int\limits^\Delta_0 \mathrm{d}t \int\limits^\infty_t  p(z) \;\mathrm{d}z\\
\Gamma_{sg}=  \frac{1}{\langle t \rangle}\int\limits^\Delta_0 \mathrm{d}t \int\limits^\infty_t  \mathrm{d}z \int\limits^\infty_z  p(x) \;\mathrm{d}x\\
\end{array}
\right.
\end{equation}
Assuming existence of second moments for latch holding time distribution we can proceed further. It is possible to change the integration orders  using:
\begin{displaymath} 
 \int\limits^\infty_t  \mathrm{d}z \int\limits^\infty_z  p(x) \mathrm{d}x  = \int\limits^\infty_t  z p(z) \mathrm{d}z - t \int\limits^\infty_t  p(z) \mathrm{d}z
\end{displaymath}
Utilizing this identity twice, we arrive to the following expression: 
\begin{displaymath} 
 \Gamma_{sg}=  \frac{1}{2 \langle t \rangle}\int\limits^\Delta_0 t^2 p(t) \; \mathrm{d}t  + \frac{\Delta}{\langle t \rangle} \int\limits^\infty_\Delta (t-\frac{\Delta}{2})  p(t) \; \mathrm{d}t  
\end{displaymath}
I will focus on two regions where analytical estimations possible. To estimate the effect of spin count tuning, we need the approximate scaling rules depending on the value of ''spin efficiency'' $\sigma=$"Spin gets/Miss".

\subsection{Spin count tuning when spin efficiency is low}
The spin may be inefficient $\sigma \ll 1$. In this   low efficiency region,  the (\ref{eq:triplei}) can be rewritten in form:
\begin{equation} 
 \left\{\begin{array}{l}
\sigma= \frac{\Delta}{\langle t \rangle} - \frac{1}{\langle t \rangle} \int\limits^\Delta_0 (\Delta - t )  p(t) \; \mathrm{d}t\\
\Gamma_{sg}=  \Delta - \frac{\Delta^2 }{2 \langle t \rangle} + \frac{1}{2\langle t \rangle} \int\limits^\Delta_0 (\Delta - t)^2  p(t) \; \mathrm{d}t  \\
\end{array}
\right.
\end{equation}
\begin{wrapfigure}{r}{0.2\textwidth}
\begin{center}
\includegraphics[scale=0.35]{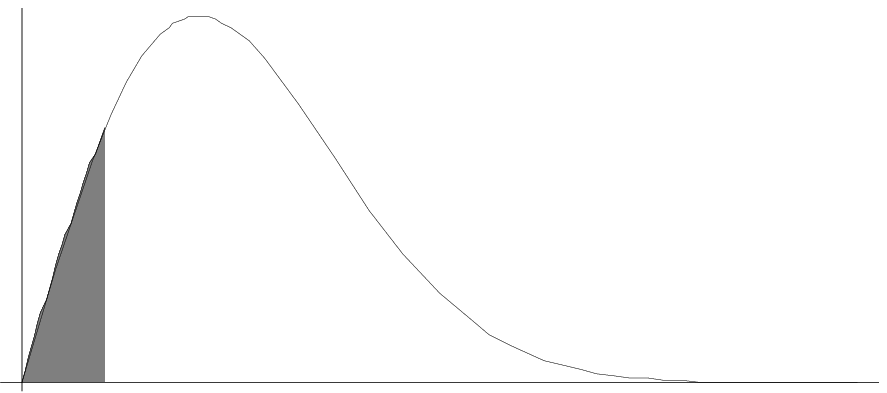}
\end{center}
\end{wrapfigure}
It is clear that such spin probes the latch holding time distribution around the origin. 

Other parts of latch holding time distribution impact spinning efficiency and CPU consumption only through the average holding time $\langle t \rangle$.
This allows to estimate how these quantities depend upon  \textbf{\_SPIN\_COUNT} ($\Delta$) change.

 If processes never release latch immediately ($p(0)=0$) then
\begin{equation} 
 \left\{\begin{array}{l}
\sigma= \frac{\Delta}{\langle t \rangle} + O(\Delta^3)\\
\Gamma_{sg}=  \Delta - \frac{\Delta^2 }{2 \langle t \rangle} +  O(\Delta^4)\\
\end{array}
\right.
\end{equation}

For Oracle performance tuning purpose we need to know what happens if we double spin count:

 {\em{}In low efficiency region doubling the spin count will double "spin efficiency"  and also double the CPU consumption.}

These estimations especially useful in the case of severe latch contention and for the another type of Oracle spinlocks --- the mutex.

\subsection{Spin count tuning when efficiency is high}

In high efficiency region, the sleep cuts off the tail of latch holding time distribution:
\begin{displaymath} 
 \left\{\begin{array}{l}
\sigma= 1 - \frac{1}{\langle t \rangle} \int\limits^\infty_\Delta (t-\Delta)  p(t) \; \mathrm{d}t\\
\Gamma_{sg}=  \frac{\langle t^2 \rangle}{2 \langle t \rangle} - \frac{1}{2\langle t \rangle} \int\limits^\infty_\Delta (t-\Delta)^2  p(t) \; \mathrm{d}t  \\
\end{array}
\right.
\end{displaymath}
Oracle normally operates in this region of small latch sleeps ratio.
Here the spin count is greater than number of instructions protected by latch $ \Delta \gg  {\langle t \rangle}$.\hfill

\begin{wrapfigure}{r}{0.2\textwidth}
\begin{center}
\includegraphics[scale=0.35]{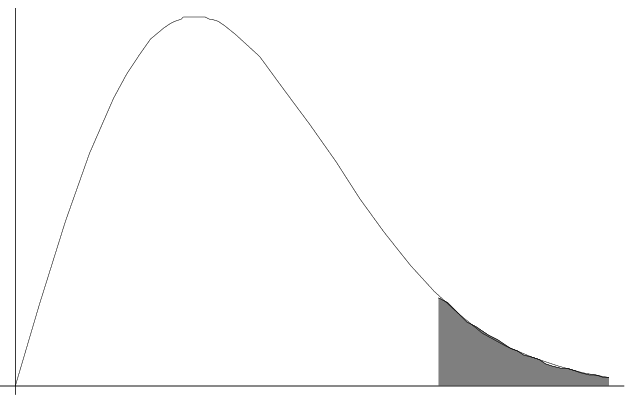}
\end{center}
\end{wrapfigure}

From the above it is clear that the spin time is bounded by both  the "residual latch holding time" and the spin count:

 \begin{displaymath} 
 \Gamma_{sg} < \min (\;\frac{\langle t^2 \rangle} {2\langle t \rangle} \;,\; \Delta)
\end{displaymath}
Sleep prevents process from waste CPU for spinning for heavy tail of latch holding time distribution

Normally latch holding time distribution has exponential tail:
\begin{displaymath} 
\left.
\begin{array}{l}
 Q(t) \sim C \exp(-t/\tau)\\
\kappa=1-\sigma \sim C \exp(-t/\tau) \\
\Gamma_{sg}\sim \frac{\langle t^2 \rangle} {2\langle t \rangle} - C \tau \exp(-t/\tau)\\
\end{array} \right.
\end{displaymath}
It is easy to see that if "sleep ratio" is small $\kappa=1-\sigma \ll 1$ then

{\em Doubling the spin count will square the ⌠sleep ratio■ coefficient.
This will only add part of order   $\kappa$    to spin CPU consumption.}

I would like to paraphrase this for Oracle performance tuning purpose as: 

{\em{}If  "sleep ratio"  for exclusive latch is 10\% than increase of spin count to 40000 may results in 10 times decrease of "latch free" wait events, and only 10\% increase of CPU consumption.}

In other words, if the spin is already efficient, it is worth to increase the spin count. This exponential law can be
 compared to Guy Harrison experimental data \cite{harrison}.

\subsection{Long distribution tails:  CPU thrashing}

Frequent origin of long latch holding time distribution tails is so-called CPU thrashing. The latch contention itself can cause CPU starvation. Processes contending for a latch also contend for CPU. Vise versa, lack of CPU power caused latch contention. 

Once CPU starves, the operating system runqueue length increases and \textbf{loadaverage} exceeds the number of CPUs. Some OS may shrink the time quanta under such conditions. As a result, latch holders may not receive enough time to release the latch. 

The latch acquirers preempt latch holders. The throughput falls because latch holders not receive CPU to complete their work. However, overall CPU consumption remains high. This seems to be metastable state,  observed while server   workload  approaches  ~100\% CPU. The KGX mutexes are even more prone to this transition.

Due to OS preemption, residual latch holding time will raise to the CPU scheduling scale -- upto milliseconds and more. Spin count tuning is useless in this case. Common advice  to prevent CPU thrashing is to tune SQL in order to reduce CPU consumption. Fixed priority OS scheduling classes also will  be  helpful. Future works will explore this phenomenon.

\section{Conclusions}

This work investigated the possibilities to diagnose and tune latches, the most commonly used Oracle spinlocks. 
Using DTrace, it explored how the contemporary latch works, its spinning-blocking strategies, corresponding parameters and statistics. The mathematical model was developed to  estimate  the effect of tuning the spin count.

The results are important for precise performance tuning of highly loaded Oracle OLTP databases.

\section{Acknowledgements}
Thanks to Professor S.V.\ Klimenko for kindly inviting me to MEDIAS 2011 conference

Thanks to RDTEX CEO I.G.\ Kunitsky for financial support.
Thanks to RDTEX Technical Support Centre Director S.P.\ Misiura for years of encouragement and support of my investigations.

%+Bibliography
\bibliographystyle{IEEEbib}

%-Bibliography
\section*{About the author}

Andrey Nikolaev is an expert at RDTEX First Line Oracle Support Center, Moscow. His contact email is \texttt{Andrey.Nikolaev@rdtex.ru}.

\end{document}